\documentclass[a4paper]{article}
\usepackage{graphicx}
\usepackage{onecolpceurws}

\usepackage{multirow}
\usepackage{makecell}
\usepackage{threeparttable}
\usepackage{cite}

\newcolumntype{P}{>{\raggedright\arraybackslash}p{3cm}}
\newcolumntype{M}{>{\raggedright\arraybackslash}m{3cm}}

\title{An Overview of Data Mining Applications in Oil and Gas Exploration:  Structural Geology and Reservoir Property-Issues}

\author{
Hamed Nikhalat Jahromi \\ LIAAD, INESC TEC \\ Portugal \\ hnj@inesctec.pt
\and
Alípio M. Jorge \\ LIAAD, INESC TEC \\ Faculty of Science, University of Porto \\ Portugal \\ amjorge@fc.up.pt
}

\institution{} 

\begin{document}
\maketitle

\begin{abstract}
Low oil prices have motivated energy executives to look into cost reduction in their supply chains more seriously.  To this end, a new technology that is experimentally considered in hydrocarbon exploration is data mining.  There are two major categories of geoscientific problems in which data mining is applied: structural geology and reservoir property-issues.  This research overviews these categories by considering a variety of interesting works in each of them.  The result is an understanding of the specific geoscientific problems studied in the literature, along with the relative data mining methods.  This way, this work tries to lay the ground for a mutual understanding on oil and gas exploration between the data miners and the geoscientists.     
\end{abstract}
\vskip 32pt

\section{Introduction}

There are two ways of increasing, or stabilizing, profit from a corporate finance perspective; either the revenues are increased or the costs are decreased.  Given the fall in oil prices since late 2014, today, cost reduction is particularly the centre of attention for many executives in the energy industry.  

Oil and gas supply chain consists of four main phases:  extraction (including exploration), refinement, transport, and distribution$\backslash$sale; for a discussion on these phases see Nikhalat-Jahromi et al. \cite{nikhalat2016future}.  In the context of cost reduction, exploration is the act of minimizing the expenses associated with finding commercial oil and gas deposits.  These expenses rise from the following main exploratory activities (in the order of being carried out):  satellite infrared and/or radar and/or microwave surveying; aerial imaging; geobotany prospecting and geochemical exploration; aerial magnetic and/or electromagnetic and/or gravity surveying; (surface) seismic surveying; and exploratory wells (generating well logs).  Kearey et al. \cite{kearey2013introduction} and Crum \cite{crum2014aerial} debate these activities thoroughly.  The key is minimizing uncertainty, i.e., increasing the accuracy with discovering omitted knowledge, in the results generated from these activities.  Note that more accurate results might decrease the number of needed activities and, therefore, aside from cost reduction, quicken the exploration and ultimately the oil extraction \cite{feblowitz2012big}. 

Data mining is a new technology applied in oil and gas exploration that can create cost reductions and bring about considerable financial gains \cite{aminzadeh2000challenges}.  It permits enhanced analysis and more effective decision-making by leveraging large volumes of data gathered in exploration \cite{abou2012data}.  In other words, data mining's potential lies in its ability to economically extract value from bulk volumes of a wide variety of data by enabling high-velocity discovery and analysis.  Feblowitz \cite{feblowitz2012big} states that (pp. 7-8):

"By applying advanced analytics, such as pattern recognition, to a more comprehensive set of data collected during seismic acquisition, geologists may be able to identify potentially productive seismic trace signatures that have been overlooked in newly acquired or archived data.  Using data from other disciplines could enhance exploration efforts.  For example, historical drilling $\dots$ data from a nearby well could help geologists and geophysicists verify their assumptions in their analysis of a field.  This becomes especially important where environmental regulations restrict new surveys."

Baaziz and Quoniam \cite{baaziz2015use} state that application of data mining in oil and gas exploration is in the experimental stage with much of the efforts focused on data-intensive computing.  Academic institutions, oil and gas companies, and business analytics service providers, are working on various applications.

This paper provides an up-to-date overview of data mining applications in oil and gas exploration, and tries to create a common understanding between the geoscientists and data miners.  It considers the two major categories of geoscientific problems in which data mining is applied and, however briefly, reviews a set of interesting works in each of them.  The categories are:  structural geology and reservoir property-issues.  In the former, data mining assists in creating and understanding of accurate images of complicated geological structures, in the reservoir and beyond, in time and depth by exploiting attributes of seismic surveying data (seismic attributes); while in the latter, this technology helps in predicting reservoir characteristics such as lithology and permeability using both seismic attributes and well logs.  

The rest of this paper is organized as follows:  Sections 2-3, respectively, cover the structural geology and reservoir property-issues.  Each section reviews the relative works chronologically and begins with the earliest research, but groups with that research all the later works which address the same geoscientific problem.  Section 4 concludes the paper and makes a few suggestions for further research.

\section{Structural Geology}

In oil and gas exploration, accurately mapping the top, bottom and lateral extent of geological structures is important as these structures might yield hydrocarbon.  This positioning, in presence of oil and gas, assists in determining the thickness of the hydrocarbon bearing (rock) layers and consequentially in deciding the economic viability of the reserve.  Furthermore, using accurate mapping, by the optimal placement of the wells, production in multi-layered reservoirs, where shale and sand layers are laminated, can be carried out from and combined into a single well, and, therefore, result in cost savings.  To the knowledge of the authors, Sriram et al. \cite{sriram1975pattern} carried out the first application of intelligent analysis techniques in oil and gas exploration.  They used non-linear mapping and a non-parametric scheme to estimate the probability density function that can produce a distribution of clusters in the data.  Sriram et al. \cite{sriram1975pattern} use their techniques for mapping laminated sand and shale sequences based on seismic data for a reservoir in the Gulf of Mexico.  Yin et al. \cite{yin1996predicting} trained a backpropagation (BP) neural network regressor for predicting the thickness of an oil deposit in China.  They correlate 10 seismic attributes, such as wavelet and Hilbert Transformation, and the known thicknesses from the exploratory wells. For introduction to and discussion on various seismic attributes see Brown \cite{brown1996seismic}.  

Later, Yu et al. \cite{yu2008dynamic} propose a dynamic all parameters adaptive BP neural network regressor for predicting the deposit's thickness by fusing genetic algorithms (GAs), simulated annealing (SA), and the BP neural network.  The fusion offsets the disadvantages of one technique by the merits of the other and eventually results in a more robust neural network.  Yu et al. \cite{yu2008dynamic} use a GA for designing the network structure.  Next, a hybrid SA-BP is implemented for finding the optimal weights and biases of the network.  These two steps are repeated recursively for various seismic attribute combinations of the training data set until the optimal network is obtained.  The method is applied to an oil reservoir in Central China.  The optimal network is made of six seismic attributes, out of 50 available ones, that are correlated to the known reservoir thickness from 15 exploratory wells.  Comparison with other methods in the literature, such a GA-SA algorithm, proves the superiority of Yu et al.'s \cite{yu2008dynamic} on this data, as it generates a smaller mean squared error (MSE) in evaluation using the test data set.         

Since faults have great significances for the storage and migration of hydrocarbon resources, fault detection is a central component of petroleum exploration.  Thus, accurately imaging faults' displacements, and determining their upthrown (the side of the fault in which rocks move upward) and downthrown sides is important.  Meldahl et al. \cite{meldahl2001identifying} present a key method for fault detection.  It comprises an iterative processing workflow using directive seismic attributes (attributes steered in a user-driven, or data-driven, direction) a BP neural network, and image processing techniques.  The network transforms all the attributes into one new attribute which indicates the probability of the fault at that position studied; it is trained on these attributes at example fault and non-fault positions.  The trained network is applied to the entire target seismic area$\backslash$section to produce a new section with high probability values at places where fault is present.  Knowledge of fault shape and orientation can be fed back to the process either to increase the detection strength or to increase the resolution of the highlighted faults.  In an example, Meldahl et al. \cite{meldahl2001identifying} fed 22 seismic attributes to a network.  These were the reference time, dip, several dip-variance, energy, similarity, and amplitude calculations.  The network made an accurate classification of the entire 3-dimentional (D) seismic section into fault and non-fault.  Next, Meldahl et al. \cite{meldahl2001identifying} applied image enhancement filters and special purpose extraction algorithms to extract fault planes in a user-controlled way.  Tingdahl and Rooji \cite{tingdahl2005semi} demonstrate the added value, in terms of fault continuity and contrast, of the combination of seismic attributes rather than a single attribute in Meldahl et al.'s \cite{meldahl2001identifying} method.  Zheng et al. \cite{zheng2014multi} apply this method successfully to a reservoir with a complex fault system in south Texas.  Results reconfirm the method's ability in suppressing the surrounding noise and highlighting the faults.  Meldahl et al.'s \cite{meldahl2001identifying} method is instrumental in both prospect evaluations (determining whether an area of exploration contains hydrocarbon in economic quantity) and research projects.  In industry, it is used as a risk reduction tool and for ranking the prospects.

\section{Reservoir Property-Issues}

The earliest discoveries of oil and gas deposits were based on finding structural traps.  But soon a different type of trap became important, stratigraphic.  Marr \cite{marr1971seismic} defines the difference between the two as:

"Structural trap is one where $\dots$ [impermeable] rocks have been folded and possibly faulted to produce a $\dots$ closure $\dots$ that is capable of trapping hydrocarbons in [below] porous and [permeable layers] $\dots$", while stratigraphic traps accumulate oil due to changes of rock character rather than faulting or folding; (permeability is the character of rock that is an indication of the ability for fluids to flow through it; porosity is the percentage of void space in the rock; and the term stratigraphy refers to the study of rocks and their variations).  For further discussion on traps see Marr \cite{marr1971seismic}.  

In finding stratigraphic traps one must go beyond structure and deduce possible lithologies (grossly meaning identifying rock layers in the subsurface) and probable presence of oil and gas.  Horizontal lithologic analysis is known as lithofacies analysis.  Justice et al. \cite{justice1985multidimensional} provided system-supported human interaction with access to automated pattern recognition procedures for comprehending rock variations.  They cluster the points of the seismic section, considering the seismic attributes, resulting in groups of lithofacies in the section.  ITERATE, a technique that sorts data and uses an iterative redistribution operator in making clusters, was proposed by Biswas et al. \cite{biswas1998iterate} as a method for clustering lithofacies in seismic sections.  To quantify the quality of the clusters generated, two probabilistic measures are introduced:  cohesion (intra-cluster similarity), and distinctness (inter-class dissimilarity).  Cohesion is measured as the increased predictability of each feature value of the objects in the data set given the assigned cluster structure.  Distinctness is quantified using a probability match measure.  Unlike deterministic evaluation measure like MSE, measures in Biswas et al.'s (1998) work are probabilistic, since ITERATE's cluster definitions are probabilistic. 

Later, Chandra et al. \cite{chandra2003lithostratigraphic} presented an application of unsupervised self-organizing maps (SOM) neural networks in lithofacies analysis.  Their clustering of the lithofacies for a reservoir in India is carried out in two steps:  first, the shape of the traces within the horizontal interval of the 3-D seismic section of the field studied are analysed by an SOM which constructs a series of model traces that best represent the diversity of the shapes observed.  Once these model traces are available, in the second step, each trace in the interval is compared to all the model traces and is assigned to the one with which it scores the best correlation.  The result is a coloured seismic section with similar lithofacies coloured alike.  Olorunsola et al. \cite{olorunsola2016multiattribute} use the Generative Topographic Mapping (GTM) technique for clustering lithofacies of a complex granite wash formation in Texas Panhandle, United States.  GTM, a non-linear projection technique, provides a probabilistic representation of the data-vectors in a corresponding lower latent dimensional space.  The features that describe the data of the 3-D seismic section in Olorunsola et al.'s \cite{olorunsola2016multiattribute} method are eight geometric, textural, and inversion seismic attributes:  coherence, coherent energy, curvature, peak frequency, gray-level co-occurrence matrices (GLCM) entropy and heterogeneity, p-impedance, and reflector convergence. 

Finally, two clustering techniques were used by Moraes et al. \cite{moraes2006cluster} for vertical lithologic analysis in a Brazilian reservoir:  fuzzy c-means, and SOM.  The authors extract three attributes (amplitude, cosine phase, and Hilbert Transformation) from the points of the 3-D seismic section of the field.  Both algorithms were run for 2-10 partitions to identify the best number of vertical lithologic groups in the seismic section and found three as the optimal value.  The quality of the groups made is assessed using PBM (acronym constituted of the initials of the names of its inventors Pakhira, Bandyopadhyay and Maulik), a cluster validation index – for discussion on PBM see Pakhira et al. \cite{pakhira2004validity}.  The values of the PBM were largely similar for the fuzzy c-means and SOM techniques for each group number (2-10) tested, indicating that on this data these techniques have comparable quality.

Xiao and Zhu \cite{baiwen1990fuzzy} analysed the fuzzy relations of, collectively; the amplitude, phase, frequency, structure curvature, interval velocity, apparent polarity, and the thickness of low velocity layer; with the nature of the hydrocarbon deposit.  They established the membership functions of these seismic attributes, classifying high-, low-, and non-productive reserves, thus properly placing the wells in the field and minimizing the number of dry (without any hydrocarbon recovery) and low-productive ones.  Yao et al. \cite{yao2004hydrocarbon} propose the feature expansion and selection in classification of high-, low-, and non-productive deposits.  It is applied in two hydrocarbon prospects in China.  Here, explicit nonlinear functions such as spline, radial basis function (RBF), and polynomial are used for feature expansion.  Features are extracted from the seismic sections of the prospects.  Expanded features tend to be more linearly related with the classification result intended.  Next, a linear support vector machine (SVM) is used for feature selection where useless attributes are removed.  An algorithm for recursive feature elimination (RFE) has been developed which trains the linear SVM and computes the weights for different attributes.  The features with smaller weights, i.e., those contributing less information, are removed and the process is repeated until no useless attribute remains.  Finally, a linear SVM is trained with selected seismic features that classifies the deposits.

Permeability estimation in exploratory wells is an important issue in determining the production potential of the deposit and therefore its economic feasibility.  Mohaghegh et al. \cite{mohaghegh1994methodological} proposed a method for assessing permeability by analysing three well logs, including gamma ray, bulk density, and deep induction log response.  For introduction to and debate on well logs see Boyer and Mari \cite{boyer1997seismic}.  They trained, using permeability of core samples from several exploratory wells of a hydrocarbon deposit in West Virginia, a BP neural network regressor.  But before that a general regression neural network is used for identifying the optimum network design to be used in the BP network.  Wong et al. \cite{wong2005reservoir} modelled estimation of permeability based on eight well logs (gamma ray, deep resistivity, shallow resistivity, flushed zone resistivity, bulk density, neutron porosity, photoelectric factor, and sonic travel time) as an SVM regression problem; permeability and the relative well logs are available for a few hundred data points for an exploratory well in a reservoir offshore Western Australia.  The regression problem turns to the optimization of a convex formulation.  Wong et al. \cite{wong2005reservoir} optimized this formulation by applying Lagrange multipliers.  The SVM regressor performs better than Mohaghegh et al.'s \cite{mohaghegh1994methodological} method in this case in that it generates a lower MSE in evaluation.
 
In an interesting project, the Correlations Company \cite{weber2001data} was commissioned by the US government to perform data mining and determine the effectiveness of water injection in enhancing hydrocarbon recovery in an oil well in Nebraska.  The effectiveness was defined as the primary to secondary (after water injection) oil extraction (P/S) ratio. The company used fuzzy logic for ranking and determining the important data attributes, out of 30 available ones from previous water injection experiences in other wells in the same region, in estimating the ratio.  Six attributes were identified as central:  water to oil ratio in the crude oil extracted from the well, gas to oil ratio, acreage dedicated to the well in the region, average porosity in the deposit beneath, average permeability, and the initial well pressure.  A neural network regressor was trained on the correlation between these six and the P/S ratio of the previous water injection experiences.  Next this network was used for prediction of the injection result in the well under question.   

Estimation of shear wave velocity in wells is a key subject which is necessary in petrophysical studies of oil and gas deposits; (shear wave is a seismic body wave that shakes the ground perpendicular to the direction the wave is moving, and petrophysics is the study of the physical and chemical properties of rocks and their contained fluids).  For many wells this velocity is not measured appropriately due to broken instruments, bad well conditions, and difficulty in accessing the equipment; or further not measured at all for cost reasons.  Eskandari et al. \cite{eskandari2004application} suggest two methods for determining the shear wave velocity in wells from well logs:  a multi regression method and a BP neural network.  The authors had access to the shear wave velocity of core samples for a carbonate oil field in the Zagros Basin, South of Iran.  Using the well logs of the carbonate oil field; compressional wave velocity, bulk density, and neutron porosity; and; neutron porosity, bulk density, transit time, gamma ray, deep laterolog, and X and Y coordinates; respectively, for the multi regression and the neural network methods, the regression is carried out and the network is trained.  Evaluation shows that the neural network here predicts the shear wave velocity slightly better, i.e., generates a better correlation coefficient than the multi regression method.  Later, a neural network with Bayesian regularization (NNBR) along with stratigraphic constraints was recommended by Srisutthiyakorn \cite{srisutthiyakorn2012redefining} for prediction of the shear wave velocity in wells from well logs.  Shear wave velocity is available for many data points from nearly 30 exploratory wells in an oil prospect onshore Thailand.  The author trains the NNBR with three well logs (gamma ray, neutron porosity, and resistivity) as attributes.  In comparison to the conventional BP neural networks, NNBR has a smaller weight and bias leading to a smoother output.  Srisutthiyakorn \cite{srisutthiyakorn2012redefining} imposes stratigraphic constraints by separating the data into geologic formations so the NNBR is trained and predicts velocities that share the same geologic settings.  This results in faster training times.  The NNBR outperforms Eskandari et al.'s \cite{eskandari2004application} multi regression method in the Thai prospect.

\section{Conclusions and Further Research}

In this paper an overview of data mining applications in oil and gas exploration was carried out.  Two major categories of geoscientific problems, which have been the focus of data mining efforts, are studied:  structural geology and reservoir property-issues.  Table 1 presents these categories and provides the key information deduced on them.

\begin{table}[htb]
\centering
\caption{Summary of the problem categories with the key information deduced}
\label{my-label}
\begin{threeparttable}
\begin{tabular} { | c | l | c | c | l | }
\hline
Category & \multicolumn{1}{c|}{Problem} & \makecell{General Data\\ Mining Method } & \makecell{Type of \\ Attributes} & \multicolumn{1}{c|}{Algorithm(s)} \\
\hline \multirow{3}{*}{\makecell{ \vphantom{}\\ Structural \\ Geology}} 
& \makecell[l]{Mapping, Laminated Sand \\ and Shale Sequences} & Clustering & \multirow{5}{*}{\makecell{\vphantom{}\\ \vphantom{}\\ \vphantom{}\\ Seismic}} & \makecell[l]{Non-Linear Mapping and\\ a Non-Parametric Scheme} \\ 
\cline{2-3} \cline{5-5} & \makecell[l]{Predicting Reservoir's \\ Thickness} & Regression & & \multirow{2}{*}{BP Neural Network} \\ 
\cline{2-3} & Fault Detection & Classification & & \\ 
\cline{1-3} \cline{5-5} \multirow{5}{*}{\makecell{ \vphantom{}\\ \vphantom{}\\ \vphantom{}\\ \vphantom{}\\ Reservoir \\Property-Issues}} 
& Lithologic Analysis & Clustering & & \makecell[l]{ITERATE; SOM Neural \\ Network; GTM; C-Means} \\ 
\cline{2-3} \cline{5-5} & \makecell[l]{Predicting Hydrocarbon \\ Deposits} & Classification & & SVM \\ 
\cline{2-5} & \makecell[l]{Determining the \\ Permeability in Wells} & \multirow{3}{*}{\makecell{ \vphantom{}\\ \vphantom{}\\  Regression}} & \multirow{3}{*}{\makecell{\vphantom{}\\ \vphantom{}\\ Well Log\tnote{*}}} & BP Neural Network; SVM \\ 
\cline{2-2} \cline{5-5} & \makecell[l]{Estimating the Secondary \\ Recovery Ratio} & & & Neural Network \\ 
\cline{2-2} \cline{5-5} & \makecell[l]{Predicting the Shear \\ Wave Velocity in Wells} & & & \makecell[l]{Multi Regression; BP \\ Neural Network; NNBR} \\ 
\hline
\end{tabular}
\begin{tablenotes}\footnotesize
\item[*] In case of "Estimating the Secondary Recovery Ratio", in addition to well logs, a few uncommon attributes are used as well.  For details see Section 3.
\end{tablenotes}
\end{threeparttable}
\end{table}

There are two ways of expanding this paper:  

(i)  There is another category of geoscientific problems in which data mining is applied.  Here, an array of ancillary data which provide high-level information in hydrocarbon exploration are covered.  Remote sensing (including satellite infrared, radar, and microwave surveying; and aerial magnetic, electromagnetic, and gravity surveying) is the most important topic in this category.  Although there are cases of data mining application to geobotany prospecting and geochemical exploration as well. In remote sensing, swaths of earth are covered, resulting in an understanding of the geological megastructures, with the aim of narrowing down to locations with the potential for hydrocarbon.  After that further surveying in these places provides precise details on the subsurface formations.  This category of problems, although less popular in academia than those two considered in this paper, is interesting to be overviewed too.

(ii)  This overview given its novel way of categorizing data mining applications in oil and gas exploration is worthy of being expanded to a comprehensive review.  Such a review will be informative to both geoscientists and data miners and deepen their mutual understanding. 
\\


\bibliographystyle{alpha}
\bibliography{ref}

\end{document}